# The role of resident electrons in manifestation of a spin polarization memory effect in Mn delta-doped GaAs heterostructures

M.V. Dorokhin[1,a], M.V. Ved'[1], P.B. Demina[1], D.V. Khomitsky[1], K.S. Kabaev[1], M.A.G. Balanta[2], F. Iikawa[2], B.N. Zvonkov[1], N.V. Dikareva[1]

[1]Lobachevsky State University of Nizhni Novgorod, 603950 Nizhni Novgorod, Russia

[2]Instituto de Física "Gleb Wataghin," Universidade Estadual de Campinas, 13083-859 Campinas, SP, Brazil

[a]E-mail: dorokhin@nifti.unn.ru

The influence of magnetic materials on the spin-dependent properties of light-emitting semiconductor structures is one of the central problems in spintronics and photonics, which if being properly solved may give rise to a number of promising applications in new generation of light-emitting devices [1-4]. One interesting manifestation of magnetic inclusions effect on LEDs operation is the interaction between spins of photogenerated carriers in the InGaAs/GaAs quantum well (QW) heterostructures and a ferromagnetic manganese monolayer located in the immediate vicinity of the QW. A number of techniques was used to demonstrate that InGaAs quantum well provides circularly polarized photoluminescence (PL) emission upon the magnetization of ferromagnetic δ<Mn> in GaAs [5-14]. The latter type of the generation of circularly polarized emission and corresponding spin polarization in InGaAs quantum well uses neither spin injection effect (see for ex. [2]) nor optical spin pumping (as in [15]) and thus is free of drawbacks attributive for each of the mentioned methods. The investigation of the exact mechanism of manganese magnetic effect on a spin polarization in a QW is a subject of interest which has been continuously explored within a decade [5-14]. Despite the long-term investigation, no general agreement has been reached on this matter yet. Earlier [7,8] it was supposed that p-d exchange interaction between the Mn ions and holes in InGaAs quantum well leads to a spin polarization of the latter accompanied with a circular polarized emission. This supposition was disputed because of a large spatial separation between a QW and δ<Mn> (up to 10 nm) exceeding the characteristic p-d interaction length for the holes. Moreover, a number of experiments studying the circular polarized PL dynamics were explained in terms of an electron spin polarization in the QW due to a spin-dependent escape from the quantum well to Mn-related defect states in the δ<Mn> surrounding [9,10]. Some later experiments have provided new data disagreeing with the conclusions made in [9-10]. These are the inversion of circular polarization sign with the variation of Mn-QW spatial separation [12] and the increase of Larmor precession frequency upon magnetizing δ<Mn> with circularly polarized light pulse [13,14]. Papers [12-14] have suggested the Mn interaction with holes in the quantum well, however the exact interaction mechanism remains a subject of discussion and its analysis requires additional experimental data.

In the present paper we provide such kind of data concerning the magnetic interactions in InGaAs/GaAs/δ<Mn> systems and specify some tools for efficient revealing of these interactions. To investigate the above system the pump-probe technique was applied. The pump pulse of circularly polarized laser emission was used to polarize Mn spins via the interaction with spin-polarized photoexcited holes [13]. The probe pulse with variable circular polarization was used to investigate the polarization state of Mn and the efficiency of Mn-hole interaction. The results obtained have shown that the resident electrons localized in the QW due to the donor doping of GaAs barrier play the dominating role in the observation of optical effects related with Mn-hole interaction. A generation-recombination balance model of the photoluminescence kinetics is derived, the model agrees with experimental data and confirms the role of resident electron concentration in obtaining the polarization dependencies in various samples.

**Experimental details**

The structures for investigation consisted of InGaAs quantum wells with a thin δ<Mn> layer introduced into a GaAs barrier as shown in Fig.1. The variable technological parameter in batch under investigation was the spatial separation between InGaAs and δ<Mn> (spacer layer thickness) which ranged between 2 and 8 nm. The samples were grown on n-GaAs (100) substrates using a hybrid system combining metal-organic chemical vapor deposition (MOCVD) and pulsed-laser ablation (PLA). First, n-GaAs buffer layer (n≈$2\times10^{16}$ cm$^{-3}$), the In$_{0.16}$Ga$_{0.84}$As quantum well (10 nm) and undoped GaAs spacer layer ($d_s$=2,4,6,8 nm) were grown by MOCVD at high temperature (~650 °C). The precursors were trimethylgallium, trimethylindium and arsine, doping was carried out using laser sputtering of Si solid target [7]. On the second stage, we have used a Q-switched YAG: Nd laser ablation system with Mn and GaAs targets for growing the Mn delta-doping layer and the GaAs capping layer ($d_c$=40 nm) respectively, the process temperature in this case was 400 °C. The entire growth process was performed in the same reactor. Further details of the growth can be found in [7,8,12].

Time-resolved photoluminescence (PL) measurements were performed using a fs Ti:Sa laser and a streak-camera system (time resolution ~50 ps). The laser wavelength was tuned for resonant QW excitation. In the present paper the two-beam excitation scheme was used as in [13]. The pump circularly polarized beam was used to excite the system and a probe beam delayed by Δ$t$ was used for probing the resulting spin states in the structure. The right- (σ +) and

left- (σ −) circular-polarized components of the excitation beams and the optical emission were selected with appropriated optics. The circular polarization of each beam can be selected independently. The time delay Δ$t$ between the pulses from the two beams was controlled by changing the optical path of one of the beams. The excitation scheme and the example of sample response are presented at Fig.1a and Fig.1b correspondingly.

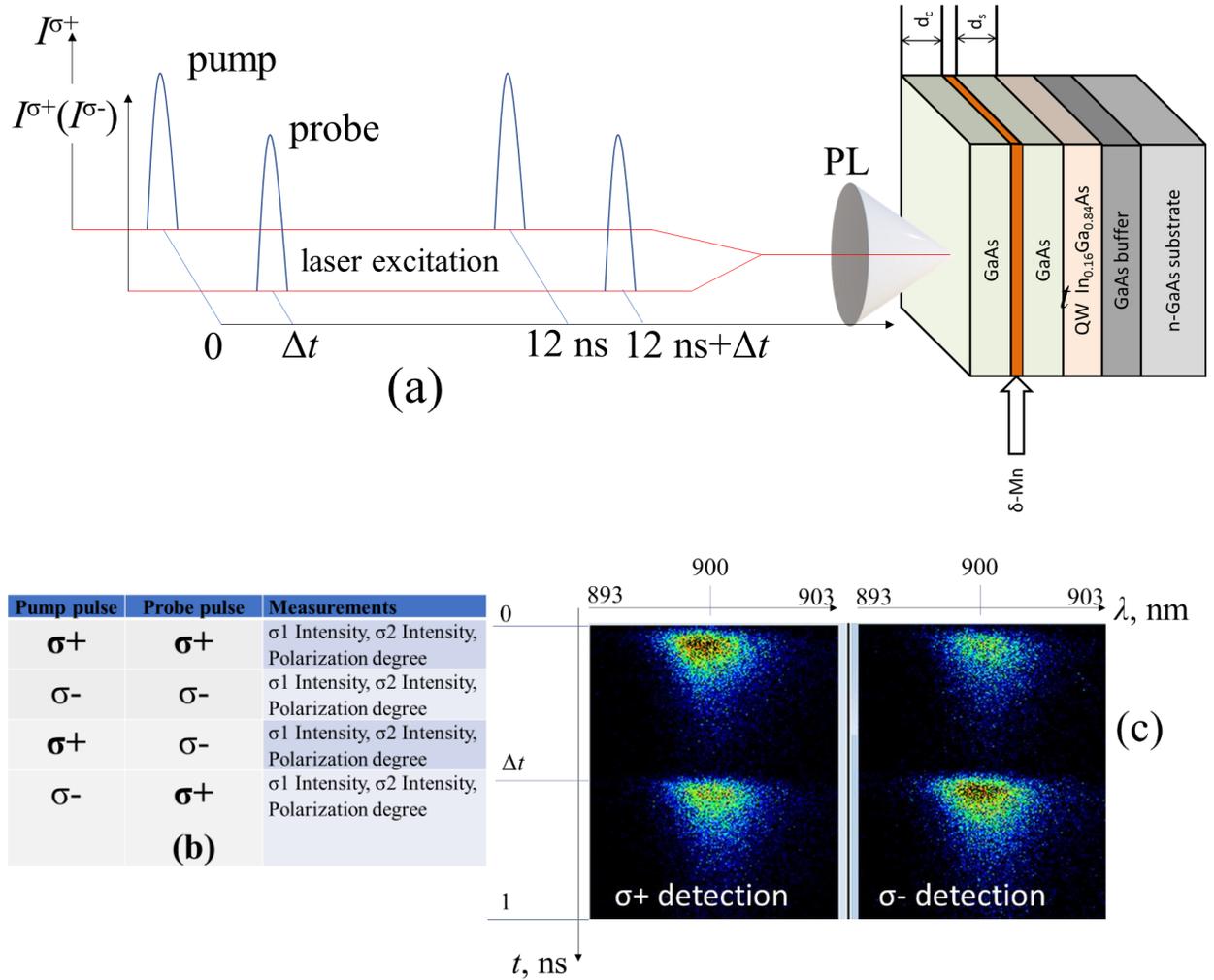

Fig.1. (a) The scheme of the sample excitation by pump and probe pulses shifted in time by Δ$t$. The scheme of the investigated sample is shown at the left; (b) the scheme of sample excitation by laser pulses and detection (c) typical time-resolved PL results from sample #4 using two excitation beams with opposite circular-polarizations. The time delay between the pulses is Δ$t$ = 0.5 ns, as shown by the schematic representation at Fig.(a). The streak camera images correspond to the σ + and σ − components of the PL emission.

From now on we refer to the first pulses of the beam that arrive first and at a time of Δ$t$ prior to the second pulse as the pump pulses. Those that arrive at the time of Δ$t$ after the first pulse will be referred as probe pulses. The results presented here correspond to the condition where the pump pulses are σ $^{+}$ polarized, and the following pulses from the second beam are σ $^{-}$ polarized. Measurements with opposite polarizations were also performed and gave equivalent results.

The degree of polarization of the PL emission is defined as:

$$Pol = (I^{\sigma+} - I^{\sigma-})/(I^{\sigma+} + I^{\sigma-}) \qquad (1)$$

where $I^{\sigma+/\sigma-}$ is the intensity of the $\sigma+/-$ emission component.

**Results and discussion**

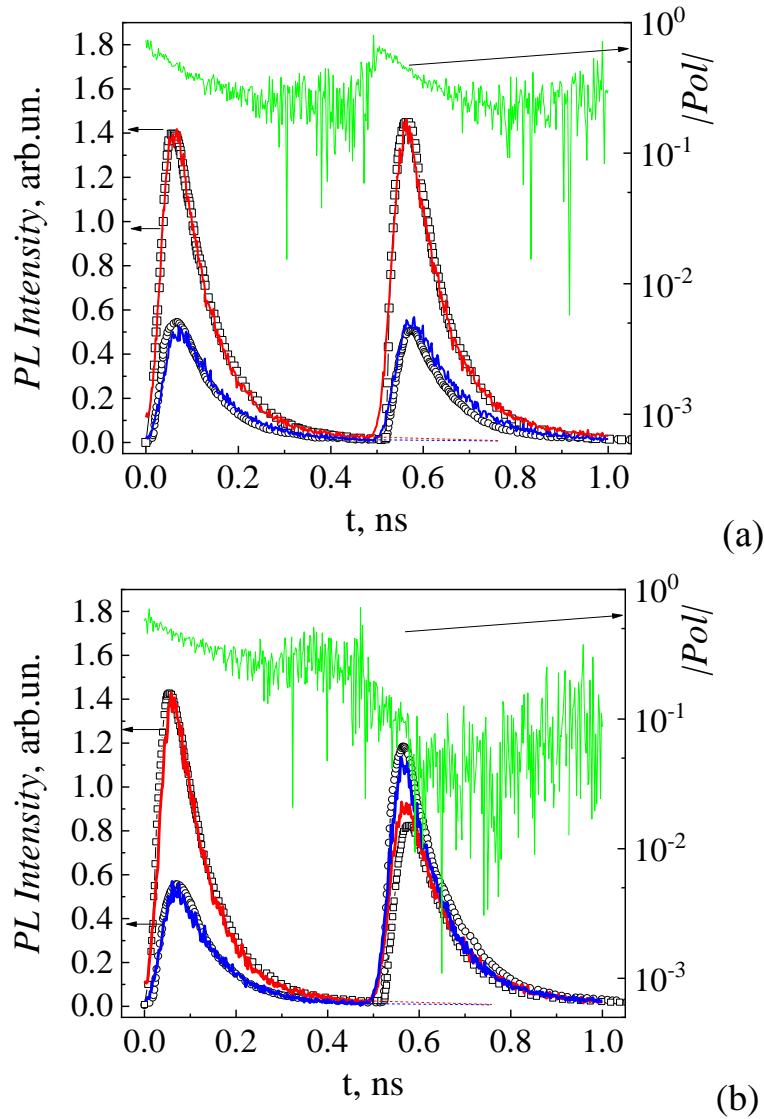

Fig.2. Time dependencies of PL intensity (red and blue curves) and absolute values of circular polarization degree (green curve) calculated by (1), for sample #4 with $d_s$=8 nm recorded in the modes of $\sigma^+$-$\sigma^+$ (Fig.2a) and $\sigma^+$-$\sigma^-$ (Fig.2b) excitations at the temperature of 7 K. Red curves correspond to $\sigma^+$ circular polarization, blue curves correspond to $\sigma^-$ circular polarization. The point plots show the intensity values calculated using a spin-dependent generation-recombination kinetic balance model (section 3), squares correspond to calculated $\sigma^+$ polarized intensities, circles correspond to $\sigma^-$ ones. Dashed curves show the exponential decay of pump PL intensities which are to be subtracted from the probe PL intensity for correct calculation of polarization of the probe PL.

The values obtained are in a good agreement with earlier results on similar samples with respect to a strong dependence of these parameters on a GaAs spacer layer thickness [13].

Table 1. The dynamic parameters of sample's photoluminescence derived from the experimental results of intensity and polarization decays.

| Sample number | $d_s$, nm | PL decay time ($\tau$), ps | Electron spin lifetime ($1/\gamma_e$), ps |
|---|---|---|---|
| #1 | 2 | 22±5 | 230±20 |
| #2 | 4 | 46±5 | 160±20 |
| #3 | 6 | 65±5 | 270±20 |
| #4 | 8 | 85±5 | 670±20 |

The probe pulse is either $\sigma^+$ or $\sigma^-$ polarized and the polarization sign of corresponding PL emission follows the polarization of laser pulse. However, the absolute value of a polarization degree for $\sigma^+$-$\sigma^-$ excitation case is significantly lower than that of $\sigma^+$-$\sigma^+$ excitation (the *Pol* value for the latter case in very close to pump emission polarization). The experimental *Pol* values at Figure 2 (curves) is influenced by the pump pulse emission. If the polarization correction is performed (by subtracting the exponential decay of the 1$^{st}$ PL pulse from the intensity of the second one [13]) there is about 40 % difference between *Pol* values for $\sigma^+$-$\sigma^+$ and $\sigma^+$-$\sigma^-$ excitation cases. According to [13] such difference is due to effect of manganese polarization on the photogenerated carriers in the quantum well. The interaction of photoexcited spin polarized holes with Mn leads to a spin polarization of the latter at the first stage of the process [13,14]. The characteristic interaction time is very short and therefore cannot be revealed within our experimental techniques. However, Mn spin lifetime is rather long and thus it can influence the polarization dynamics even after the delay of $\Delta t$ between pump and probe pulses. Fig.3 shows the dynamics of the polarization difference value ($\Delta P$) calculated as:

$$\Delta P(t) = \mathrm{abs}(|Pol^{\sigma^+ - \sigma^+}(t)| - |Pol^{\sigma^+ - \sigma^-}(t)|) \qquad (2)$$

where $Pol^{\sigma^+ - \sigma^+}$ is a polarization degree for $\sigma^+$ - $\sigma^+$ excitation and $Pol^{\sigma^+ - \sigma^-}$ is a polarization degree for $\sigma^+$ - $\sigma^-$ excitation scheme.

Since we have selected the $\sigma^+$ polarization of pump in both cases the $\Delta P$ value for the 1$^{st}$ PL pulse is zero with respect to experimental error. The probe laser pulse, on the contrary is of different polarizations, which gives rise to nonzero values of $\Delta P$ for the 2$^{nd}$ PL pulse. We note that there is only a short time period of rather high PL intensity during which $\Delta P$ value can be defined without a huge experimental error. As the intensity decreases the experimental error value raises significantly. For that reason, we will not discuss here the time dependence of $\Delta P$ and confine ourselves with the values averaged over the time interval of ☐100 ps (when the high

PL intensity is preserved). Even with such limitation the $\Delta P$ is a very important parameter defining the behavior of the system.

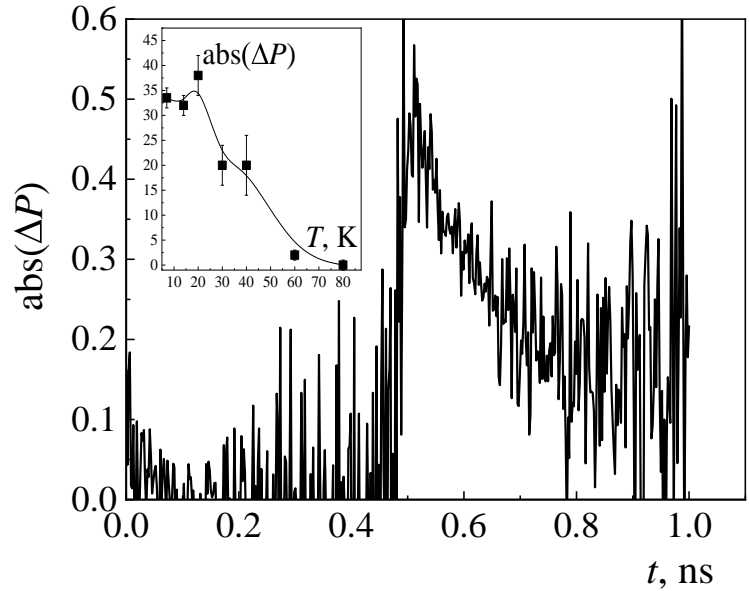

Fig.3. Time dependence of $\Delta P$ value for sample #4 calculated as in (2) at 7 K. The insert shows the temperature dependence of $\Delta P$ averaged over a time between 0.5 and 0.6 ns.

The Insert to Fig.3 shows the temperature dependence of $\Delta P$-effect. At the temperature range between 7-20 K the $\Delta P$ value does not change above the limits of an experimental error. Above 20 K a rapid decrease of $\Delta P$ is observed. However detectable $\Delta P$ effect was recorded even at 70 K which is above the Curie temperature of δ<Mn> (30-40 K [7,8]).

Figure 4 shows the dependence of $\Delta P$ on the delay time for all investigated samples. With the increase of the delay the values $\Delta P$ decrease which should take place due to a spin relaxation of Mn. The estimated Mn spin relaxation time is about 5 ns which although rather high yet can be visualized within the explored time windows. The $\Delta P$ decrease for small delays for the sample #4 cannot be unambiguously confirmed because of the huge experimental errors raising from the necessity to subtract the intensity decays from the 1[st] pulse. Taking the experimental error into account we can only state that $\Delta P$ does not change significantly for the $\Delta t$ range from 300 to 500 ps. The characteristic decay time of the $\Delta P$-effect can be estimated for samples #3 and #4 as approximately 5 ns. For the samples #1 and #2 this value cannot be estimated due to a relatively big experimental error and small initial value of the effect.

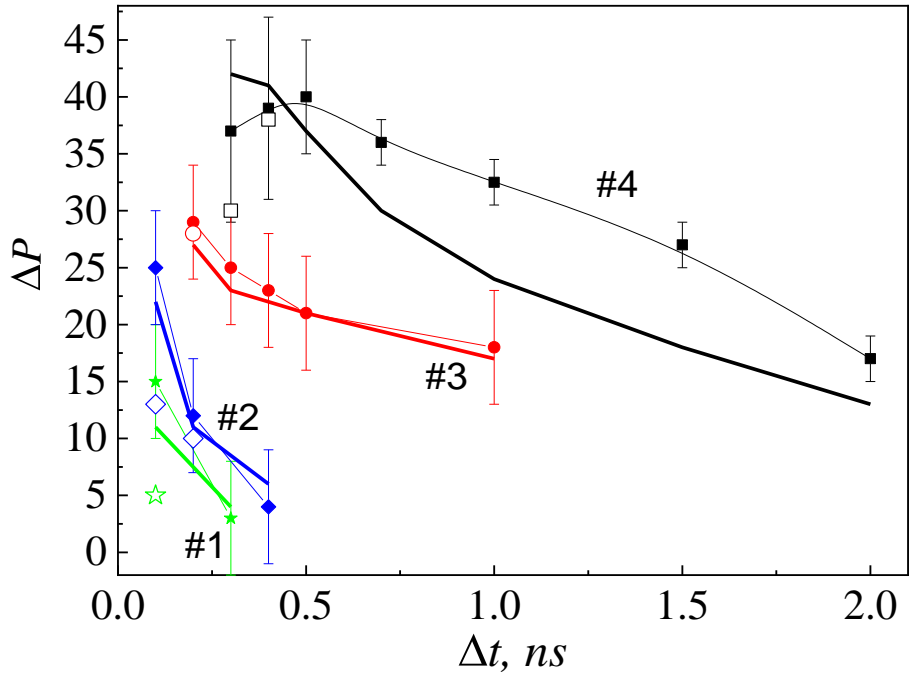

Fig.4. The dependence of averaged $\Delta P$ on the delay time between two pulses ($\Delta t$) for all investigated samples. Solid dots correspond to experimental measurements, open dots correspond to the values calculated after subtracting the exponential decay of the first pulse from the intensity of the second one. Thick solid lines correspond to the $\Delta P$ values calculated using a kinetic balance model presented in the Discussion section of the paper.

The most surprising result is a decrease of $\Delta P$ value with the decrease of the spatial separation between δ<Mn> and a quantum well. Indeed, the amplitude of various Mn-hole interaction effects usually increases as $d_s$ decreases [6-12]. Nevertheless, the greatest $\Delta P$ effect was detected for the sample with the highest $d_s$ among the entire set. The latter result does not agree with the one obtained in [13] for similar samples, the reason for such a disagreement will be discussed in the next section of the paper.

**Kinetic balance model**

Let us now switch to a discussion of the experimental results. First, we will briefly discuss the qualitative model of dynamic spin polarization in the investigated samples, then we will present the results of theoretical calculations which were performed by solving the balance equations. Consider a system of energy levels in a strained quantum well with enclosed δ<Mn>layer shown in Fig.5. The distinguishing feature of our system is that GaAs substrate and the buffer layer are donor-doped (unlike [13]). The quantum well is located in the space charge region of the n-GaAs/(Ga,Mn)As p-n junction (as Mn is an acceptor impurity). The type and concentration of carriers in the quantum well thus depend on its position relative to the boundaries of the p- and n-regions. We assume that in a steady state condition, conduction electrons from the n-GaAs buffer layer are localized in a QW. The greater separation between the QW and delta layer, the higher is the concentration of these "resident" electrons is. For the sample #4 with $d_s$ = 8 nm, the electron concentration in the QW is maximum among all structures, and for sample #1 with $d_s$ = 2 nm, it is minimum (close to zero).

1. Prior to the circularly polarized pump pulse arrival (i.e., without any external fields) the polarization of resident electrons is zero i.e. the concentrations of resident electrons with the spins of +1/2 and -1/2 are equal (Fig.5a).

2. As a pump pulse arrives the electron-hole pairs are generated in the QW, as is shown in Fig.5b. The photoexcited holes and electrons are spin polarized with respect to selection rules. The polarized photoexcited electrons with the concentration of $N$ are added to unpolarized resident electrons in the QW.

3. Within the time of less than 100 ps the spin orientation of Mn ions occurs due to interaction with holes (Fig.5c). This process and its characteristic timescale were experimentally confirmed for similar systems in [13,14].

4. Within the time of about 100 ps fast hole spin relaxation process provides the redistribution of heavy hole spins into some equilibrium state. We note that in general 3/2-spin and -3/2-spin concentrations are not equal to one another due to a Mn spin polarization influence of heavy holes (Fig.5d).

5. At the next stage, the radiative recombination of spin-polarized photoexcited carriers starts to play role in electron-hole dynamics (Fig.5e,f). Radiative transitions proceed in accordance with the selection rules and taking into account the spin polarization of electrons and holes. Since each of the spin states in a system is filled with electrons and holes (with different concentrations though), the degree of polarization is less than unity. The value of *Pol* is determined by the spin polarization of electrons and holes. As the radiative recombination proceeds, spin relaxation of electrons takes place, as a result circular polarization degree decreases with time. The spin relaxation time of electrons and the radiative lifetime are the same order of magnitude; therefore, the processes of recombination and spin relaxation of electrons are clearly visible in the recombination dynamics.

6. Prior to a probe pulse arrival, the carrier distribution between the energy levels is similar to the one shown at Fig.5a with the only exception of Mn being spin polarized since the Mn spin decaying time is greater than the delay times used [14]. The QW levels are filled with equilibrium electrons with equal or nearly equal concentrations of +1/2 and -1/2 spins (Fig.5g).

7. When the probe pulse polarization is σ − the spin polarization of photoexcited carriers is inverse to the one shown at Fig.5b (Fig.5h). Next process is a hole spin relaxation to the equilibrium state which is somewhat different from the one shown at Fig.5c. Due to a Mn-hole interaction the concentration of holes with a minority spin is greater than majority spin concentration (Fig.5i).

(a) – prior to pump pulse

(b) – pump pulse arrival, the pump pulse polarization here is $\sigma^+$

(c) – < 100 ps after the pump pulse, Mn is polarized by holes

(d) ≈ 100 ps after pump pulse arrival, spin relaxation of holes takes place;

(e) – radiative recombination within the radiative lifetime;

(f) the $\sigma^+$ and $\sigma^-$ intensities and $Pol$ after pump illustrating the recombination process

(g) – after recombination is over and prior to probe pulse

(h) – < 100 ps after probe pulse arrival, the probe pulse polarization here is $\sigma^-$

(i) – 100 ps after probe pulse arrival, spin relaxation of holes takes place

(j) – radiative recombination within the radiative lifetime;

Fig.5 Scheme of filling of energy levels in the strained quantum wells with respect to polarization, carrier generation by pump and probe pulses and recombination. The arrow shows the spins of a different sign. The big arrow represents the spin of manganese.

8. As a result, the spin-dependent radiative recombination conditions change. The PL polarization in this case is a result of competition between the minority and majority carrier spin transitions (the concentration of minority holes is greater than that of majority holes whereas concentration of majority electrons is greater than that of minority). Such competition leads to a decrease of circular polarization degree in absolute value (Fig.2b). The intensity of recombination of minority spin carriers in fact depends on the concentration of resident electrons, since at the initial moment of time the spin relaxation of electrons is small and only resident electrons can participate in recombination with minority holes (Fig.5j).

In the samples with higher concentration of resident electrons, the "minority" recombination intensity is also higher and the polarization degree is smaller. This explains the "inverse" dependence of the spin memory $\Delta P$-effect on the thickness of the spacer layer. With a large thickness of spacer GaAs, the concentration of resident electrons is the highest, which makes the maximum contribution to the intensity of "minority" polarized PL and, accordingly, decreases the overall circular polarization degree. As the thickness of the spacer layer decreases, the concentration of resident electrons decreases, thus decreasing the number of recombination events with a minority spin. The efficiency of the interaction between the spins of holes and manganese in the structures under study plays a secondary role. According to the numerous experimental data [12,16], the efficiency of interaction between Mn and holes weakly depends on the thickness of the GaAs spacer layer.

**Modelling**

The above processes were modeled via the time-dependent spin-resolved equations describing hole and electron generation by laser pulses, the interaction of holes with *Mn* ions in the delta layer, and the carrier recombination with emission of the circular polarized photons [13]. They include the electron concentrations $N^e_{1,2}(t)$ with spin $-1/2$ or $+1/2$ respectively, the heavy hole concentrations $N^h_{1,2}(t)$ with spin $\mp 3/2$, and the *Mn* ion concentrations $N^{Mn}_{1,2}(t)$ which gives in total six components and the corresponding system of six balance equations governing the evolution of the concentrations [17,18].

The change in concentrations of spin-polarized electrons, holes and Mn follows the balance equations:

$$\frac{dN^e_{1,2}}{dt} = A\sigma_{2,1}(t) - B \cdot \min\left(N^e_{1,2}, N^h_{2,1}\right) - \gamma_e \Delta N^e_{1,2}$$
$$\frac{dN^h_{1,2}}{dt} = A\sigma_{1,2}(t) + C \cdot f\left(N^h_{1,2}\right) \cdot \Delta N^{Mn}_{1,2} - B \cdot \min\left(N^e_{2,1}, N^h_{1,2}\right) - \gamma_h \Delta N^h_{1,2}. \qquad (3)$$
$$\frac{dN^{Mn}_{1,2}}{dt} = D \cdot f\left(N^{Mn}_{1,2}\right) \cdot \Delta N^h_{1,2} - \gamma_{Mn} \Delta N^{Mn}_{1,2}$$

The electrons and holes may be created with the absorption of incident photons, or annihilated with the emission of photons having the polarization depending on the electron and hole spin projections. The first term in equations 1 ($A\sigma_{2,1}(t)$) refers to the generation of electro-hole pairs by the laser pulses $\sigma_{2,1}(t)$ referring to $\sigma^+$ and $\sigma^-$ circular polarizations correspondingly. The laser pulses have the delta-shaped profile corresponding to the short duration $\tau_p = 0.1$ ps and are spaced from one another by the window $\Delta t$ ranging from 300 to 2000 ps. Besides the generating terms stemming from the laser pulses, the balance equations for electrons and holes include the annihilation terms describing the photoluminescence proportional to the minimum of the concentrations for the two participating components ($B \cdot \min\left(N_{i,j}^e, N_{j,i}^h\right), i,j = 1,2$). These terms take into account the selection rules for optical transitions in the quantum well.

The third group of terms describes the interaction between the holes and the spins in the *Mn* layer. These the phenomenological terms depend on the hole and *Mn* polarization (the difference in hole and *Mn* spin concentrations), they can be introduced as $C \cdot f\left(N_{1,2}^\square\right) \cdot \Delta N_{1,2}^{Mn}$ for holes and $D \cdot f\left(N_{1,2}^{Mn}\right) \cdot \Delta N_{1,2}^\square$ for Mn in the second and third pairs of equations (3) respectively. Here $C, D$ are Mn-hole interaction constants. The function $f\left(N_{1,2}^\square\right)$ describes the spin channel from which the hole spins can be taken for increasing the corresponding spin population. Consider $\frac{dN_1^h}{dt}$, the change in $N_1^h$ is only possible when there is $N_2^h \neq 0$. In this case $f\left(N_{1,2}^\square\right) = N_2^h$ and the entire term for Mn-hole interaction is $\frac{dN_1^h}{dt}(Mn - \text{hole}) = C \cdot N_2^h \cdot \Delta N_1^{Mn}$. Similar function $f\left(N_{1,2}^{Mn}\right)$ is present in the balance equation for *Mn* spins. The concentration dependencies of $f\left(N_{1,2}^\square\right)$ and $f\left(N_{1,2}^{Mn}\right)$ create a nonlinearity in the system (3) so it is needful to use numerical methods for solving it. Then, from (3) the conservation of the total *Mn* concentration $\left(N_1^{Mn} + N_2^{Mn}\right)$ follows since $d\left(N_1^{Mn} + N_2^{Mn}\right)/dt = 0$ while the hole and electron concentrations are not conserved due to the laser pulse generation and subsequent photoluminescence.

The final group of terms in all equations describes spin relaxation with characteristic rates $\gamma_e$, $\gamma_h$, and $\gamma_{Mn}$ for the electrons, the holes, and *Mn* spins respectively. The relaxation terms are proportional to the corresponding differences in spin-resolved concentrations $\Delta N_{1,2}^e$, $\Delta N_{1,2}^\square$, and $\Delta N_{1,2}^{Mn}$. Here for brevity, we define $\Delta N_1^e = N_1^e - N_2^e$, $\Delta N_2^e = N_2^e - N_1^e = -\Delta N_1^e$, and so on.

The relaxations rates in system (3) are taken from the tables of corresponding material parameters while the constants *A, B, C, D* are mainly fitting parameters determined from the experimental data on the photoluminescence. The system (3) is accompanied by the initial conditions where we set the total concentration of *Mn* spins interacting with holes to *1* as a unit for non-dimensional concentrations so $N_1^{Mn}(0) = N_2^{Mn}(0) = 1/2$, and express the initial non-zero electron concentrations via this unit. The initial hole concentration in our samples is zero.

When the spin-resolved concentrations are found as a solution of system (3), one can write down the circular photoluminescence intensities ($I_{01}$ is the left polarization and $I_{02}$ is the right polarization)

$$I^{\sigma+} = B \cdot \min\left(N_2^e, N_1^h\right)$$
$$I^{\sigma-} = B \cdot \min\left(N_1^e, N_2^h\right). \quad (4)$$

After obtaining the intensities one gets also the polarization defined by (1). The time dependencies of the recorded intensities (4) and the polarization (1) will be the subject of our interest in the following section for two cases of laser pulse:

Parallel case 1: $\sigma^+$ pulse followed after window $\Delta t$ by the same $\sigma^+$ pulse;

Antiparallel case 2: $\sigma^+$ pulse followed after window $\Delta t$ by the oppositely polarized $\sigma^-$ pulse.

Following the experimental section, the polarization (1) for each pulse sequence is labeled as *Pol*$^p$ (parallel) and *Pol*$^a$ (antiparallel) and the corresponding polarization module difference $\Delta P$ is defined by (2). The value of $\Delta P$ reflects the influence of the holes and the *Mn* layer. The long spin relaxation time for *Mn* creates a "spin memory effect" which shifts the hole spin polarization following the sign of *Mn* spin polarization, leading to the decrease in the absolute value of polarization (1) after the second laser pulse.

One other notice considering the investigated system is the equality of polarizations after 1st and 2nd pulse of the same polarity (Fig.2a) which provides a very important condition of Mn polarization by the laser pulse. We believe that the holes generated by the pump laser pulse provide near 100 % spin polarization of interacting Mn atoms and hence the probe laser pulse of the same polarity does not lead to a significant change in Mn polarization. On the contrary, the probe laser pulse of inverse polarity can lead to reorientation of Mn spins.

**Results of modeling**

We consider the following parameters for the sample #4 where the maximum polarization difference has been achieved.

1. The initial concentrations $N_1^{Mn}(0) = N_2^{Mn}(0) = 1/2$; we choose $N^{Mn}$ as a reference unit of concentration. The resident electron concentration is $N_1^e(0) = N_2^e(0) = 2.26 \cdot 10^{-4}$, which has been estimated from the band picture of the system at 10 K. The resident hole concentration is zero ($N_1^h(0) = N_2^h(0) = 0$), since these conditions should provide the dependence shown at Fig.3.

2. 2. The relaxation rates in units of 1/ps are $\gamma_e=1.493\cdot 10^{-3}$ 1/ps, $\gamma_h=0.027$ 1/ps, as has been evaluated earlier for the similar systems [14]. From the photoluminescence experiments presented in [14] we get $\gamma_{Mn}=2.0\cdot 10^{-4}$ 1/ps.

3. The remaining parameters $A=0.0128$ 1/ps and $B=0.0125$ 1/ps define the generation and photoluminescence rates. The parameter B is defined via the experimentally measured photoluminescence data and is sample-dependent while the generation rate is assumed to be constant for all samples. The parameters $C$ and $D$ should be for the best fit to the experimental data. Being multiplied by the hole and Mn spin-resolved concentrations $N_{1,2}^h$ and $N_{1,2}^{Mn}$ they define in system (3) the polarizations rates for hole and Mn spins due to interaction with each other. For the given experimental conditions maximal achievable hole concentration (in units of Mn spins) reach the value $N_{1,2\ max}^h = 0.0013$ for all samples while $N_{1,2\ max}^{Mn} = 1$.

The results of modeling for the recorded intensities (3) are shown in Fig.2 for the window between the laser pulses of $\Delta t=500$ ps. Since the detector camera has a finite time resolution defined as a convolution window $\Delta t_c$ which is about 50 ps in our experiments. For that reason, calculated intensities were expressed via the camera convolution kernel $K(t-\tau)$ in the Gaussian convolution form. With the detector convolution function taken into account there is a good agreement between experimentally measured PL pulses and curves calculated upon the solution of equations in system (3). This proves that the experimentally revealed behavior of QW/δ<Mn> system can be adequately explained within our model.

Another important parameter is the total concentration of Mn spins $N_1^{Mn} + N_2^{Mn}$ actively interacting with holes ans serving as a unit of concentration. It is a natural suggestions that this number is sample-dependent since the distance $d$ between the Mn layer and the holes in QW is varied. The greater is $d$, the smaller fraction $N(d)$ of Mn spins is expected to interact efficiently. If we expect a tunnel-like overlap of Mn and hole wavefunctions, than an exponential fit of the present form may be considered:

$$N(d) = N(d_0)\mathrm{Exp}\left(-\frac{d-d_0}{d_0}\right) \qquad (5)$$

where $d_0$ and $N(d_0)$ are the reference distance and concentrations which we take for the sample #4 with $d=8$ nm. The values of both pre-set and fitting parameters for Eq.(3) are listed in Table.2 for all set of samples. It should be noted that for setting the initial electron concentration values in the last line of Table 2 we take into account the experimentally obtained absolute values of the resident electron concentration as well as the scaling function (5).

Table 2. The values of pre-set and fitting parameters for eq.(3)

| Sample # | 1 ($d_s$=2 nm) | 2 ($d_s$=4 nm) | 3 ($d_s$=6 nm) | 4 ($d_s$=8 nm) |
|---|---|---|---|---|
| A – hole and electron generation rate | 0.0128 1/ps | 0.0128 1/ps | 0.0128 1/ps | 0.0128 1/ps |
| B - radiative recombination rate | 0.0454 1/ps | 0.0212 1/ps | 0.0161 1/ps | 0.0125 1/ps |
| $C \cdot f\, N_{1,2\,max}^{h}$ – rate of hole polarization by Mn | 0.044 1/ps | 0.044 1/ps | 0.044 1/ps | 0.044 1/ps=27 ps |
| $D \cdot f\, N_{1,2\,max}^{Mn}$ – rate of Mn polarization by holes | 0.0325 1/ps | 0.0325 1/ps | 0.0325 1/ps | 0.0325 1/ps=31 ps |
| $\gamma_e$ | 0.004348 1/ps = 230 ps | 0.005882 1/ps = 170 ps | 0.003703 1/ps = 270 ps | 0.001493 1/ps=670 ps |
| $\gamma_h$ | 0.027 1/ps | 0.027 1/ps | 0.027 1/ps | 0.027 1/ps=37 ps |
| $\gamma_{Mn}$ | 0.0002 1/ps=5 ns | 0.0002 1/ps=5 ns | 0.0002 1/ps=5 ns | 0.0002 1/ps=5 ns |

The calculation procedure performed for the sample #4 with *Δt=500* ps was first repeated for different delay times and theoretical dependence of Δ*P* vs *Δt* was plotted (Fig.3, thick lines). We believe that the parameters of the equation discussed above are fundamental characteristics of system behavior and therefore should not change with a change in the delay time. For that reason, we have used all the same modelling parameters as in the 500 ps case. However, the change in the sample structure, in particular the spacer layer thickness might deviate some of the equation (3) parameters discussed in the precious section. For that reason, when performing the same calculation procedure for the rest of the batch one should re-normalize the calculation parameters.

The carrier generation rate is considered independent on the spacer layer thickness provided that In content in QW is same for all of the structures. The recombination rate as well as the rate of electron spin relaxation can be derived from experimental dependencies of the PL intensity and polarization decay as has been discussed above. Then we believe that Mn vs hole interaction constants (C and D) as well as hole and Mn spin relaxation rates weakly depend on the spacer layer thickness. The latter can be confirmed by the paper [16] in which very weak dependence of Mn-hole interaction on the spacer layer thickness was revealed.

In fact, the parameter that changes most drastically with a decrease of a spacer layer thickness is a residual electron concentration in the quantum well ($N_{1,2}^{e}(0)$). Indeed, this concentration should decrease upon shifting the QW towards acceptor-type δ<Mn>-layer. In addition, the variation of $N_{1,2}^{e}(0))$ is a corner stone of the qualitative model presented at Fig.5. We believe that small changes of the equation constant bring insignificant contribution into the Δ*P* variation as compared with the change of $N_{1,2}^{e}(0)$. With respect to above considerations, we have chosen the same fitting parameters for all investigated samples with the only exception for residual electron concentration. As a result, the Mn-hole spin interaction dependence on the distance *d* is included in our model via the resident electron's concentration scaling function (5).

The results of simulations are presented at Fig.3 (thick lines) and Fig.6. Let us first discuss the Fig.6 which shows the time dependence of Mn spin polarization in case of pumping with same polarization and different polarization of pulses. One can see that indeed pumping the electron-hole system in the QW leads to high polarization of Mn due to the interaction terms *C* and *D* (see eq.3). We also note a significant asymmetry between the cases of same and different polarizations. In the 1$^{st}$ case the probe pulse only slightly increases the polarization since the Mn subsystem is already polarized by the pump pulse. In the 2$^{nd}$ case the probe pulse switches the Mn polarization to almost zero (by the absolute value). In such case Mn-hole interaction should make smaller contribution into the resulting polarization of the PL after the probe pulse, which is confirmed by the experiment. Thus, hole polarization is indeed influenced by the *Mn* layer working in opposite to probe pulse polarization.

Since *Mn* spins are subject to spin relaxation on the characteristic time $\tau_{Mn}=1/\gamma_{Mn}$, one can expect a decrease of the "spin memory" effect when the window between the pulses is extended and the *Mn* spin system loses some of its polarization before the second pulse arrives.

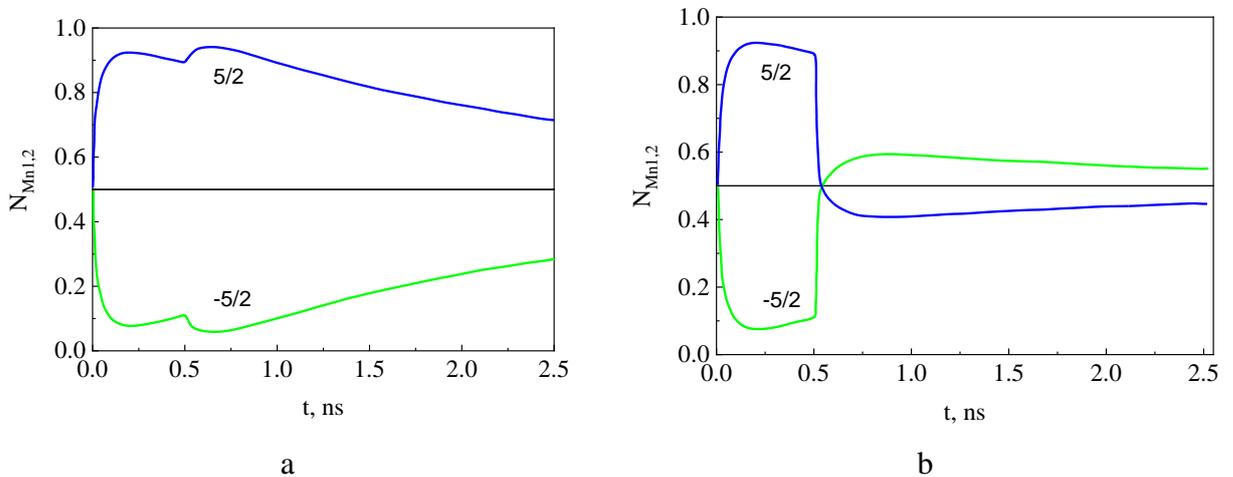

Fig.6. The calculated time dependence of concentration of Mn atoms in δ<Mn>-layer having the polarization of 5/2 and -5/2 upon the excitation of electron-hole subsystem by the pulses of $\sigma^+$ - $\sigma^+$ excitation sequence (a) and $\sigma^+$ - $\sigma^-$ excitation sequence (b).

Fig.3 (thick curves) shows the calculated dependencies of Δ*P*-effect on time delay between the pulses. The calculated dependencies are in satisfactory agreement with the experimental ones (compare dots and thick lines at Fig.3) with the only exception for sample #4. Despite some quantitative disagreement one can state that model of the photoluminescence kinetics (3) when correctly applied reflects the trends of the investigated system behavior discussed above. First, with the increase of $\Delta t$ the $\Delta P$ value decreases. The extracted relaxation time for *Mn* spins is about *5* ns which agrees with other experimental results [13,14]. Second, the decrease of resident electron concentration leads to decrease of spin memory effect. As a final note we should mention that our model takes into account the presence of pump-generated carriers in the QW at the moment of probe pulse arrival in case of short $\Delta t$ values. For that reason, the calculated values of $\Delta P$ include contribution from both Mn-affected polarization and pump pulse polarization. In order to exclude the second factor in the first approximation one is to subtract the exponential decay of pump PL intensity from the PL intensity of the second pulse. Such procedure was performed for the experimental data, the results are shown at Fig.3 as open dots.

**Conclusions**

Thus, we have investigated the spin-memory effect in the GaAs/InGaAs heterostructures with δ<Mn>-layer in GaAs barrier. The effect consists in spin polarization of Mn atoms due to interaction with photogenerated spin-polarized holes. The investigation of the effect was carried out by analyzing the polarization of the probe photoluminescence pulse in the pump-probe technique. It was shown that the circular polarization degree of probe pulse generated photoluminescence is strongly affected by the interaction of hole spins with spins of Mn atoms polarized by the pump pulse. The latter leads to decrease of circular polarization degree as compared with single pulse excitation ($\Delta P$-effect). The amplitude of $\Delta P$-effect is most strongly affected by the concentration of resident electrons in the quantum well which is believed to be due the specific compliance with selection rules for optical transition with the participation of unpolarized resident electrons. The rest of the sample's parameters including the spatial separation between δ<Mn>-layer and InGaAs quantum well ($d_s$) have a minor effect on the $\Delta P$ value which leads to a paradoxical situation of decreasing $\Delta P$-effect with the decrease of $d_s$. The proposed experimental technique consisting in creating the significant concentration of resident electrons in the QW may serve as a reliable photoluminescence method determining the strength of this effect as well as the *Mn* spin relaxation time in a particular nanostructure.

**Acknowledgements**

This work was supported by the 5-100 Competitiveness Enhancement Program.